# Trivalent ion overcharging on electrified graphene

Amanda J. Carr, Sang Soo Lee, Ahmet Uysal*
Chemical Sciences and Engineering Division, Argonne National Laboratory, Lemont, IL 60439, United States
*Corresponding Author Email: auysal@anl.gov

**Abstract**

The structure of the electrical double layer (EDL) formed near graphene in aqueous environments strongly impacts its performance for a plethora of applications, including capacitive deionization. In particular, adsorption and organization of multivalent counterions near the graphene interface can promote nonclassical behaviors of EDL including overcharging followed by co-ion adsorption. In this paper, we characterize the EDL formed near an electrified graphene interface in dilute aqueous $YCl_3$ solution using *in situ* high resolution x-ray reflectivity (also known as crystal truncation rod (CTR)) and resonant anomalous x-ray reflectivity (RAXR). These interface-specific techniques reveal the electron density profiles with molecular-scale resolution. We find that yttrium ions ($Y^{3+}$) readily adsorb to the negatively charged graphene surface to form an extended ion profile. This ion distribution resembles a classical diffuse layer but with a significantly high ion coverage, i.e., 1 $Y^{3+}$ per 11.4 ± 1.6 $Å^2$, compared to the value calculated from the capacitance measured by cyclic voltammetry (1 $Y^{3+}$ per ~240 $Å^2$). Such overcharging can be explained by co-adsorption of chloride that effectively screens the excess positive charge. The adsorbed $Y^{3+}$ profile also shows a molecular-scale gap (≥5 Å) from the top 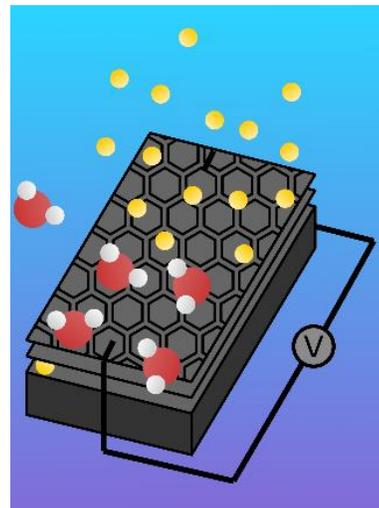 graphene surfaces, which is attributed to the presence of intervening water molecules between the adsorbents and adsorbates as well as the lack of inner-sphere surface complexation on chemically inert graphene. We also demonstrate controlled adsorption by varying the applied potential and reveal consistent $Y^{3+}$ ion position with respect to the surface and increasing cation coverage with increasing the magnitude of the negative potential. This is the first experimental description of a model graphene-aqueous system with controlled potential and provides important insights into the application of graphene-based systems for enhanced and selective ion separations.

## Introduction

A fundamental, comprehensive understanding of water and metal ion behavior near solid graphene is critical for basic science and promising graphene technology, including graphene-based separation membranes for metal extraction, ion sorption, and transport;[1] biomolecular self-assembly; and surface chemical reactions[2] within catalysis;[3] and energy storage.[4] Graphene is an ideal material for many of these applications as it is mechanically robust, lacks surface complexity, and is highly conductive.[5] In particular, cation and anion organization near the graphene interface is important for separations technologies, such as capacitive deionization (CDI).[6] These systems rely on the electrical double layer (EDL) formed near the graphene surface to remove target ions. Previous works have elucidated the importance of an uncrowded and thin EDL for improved system performance. Counterions that gather near the working electrode extend the EDL and diminish charge storage efficiency[7] and can promote overcharging via local ion screening. Despite the importance of these phenomena on overall system efficiency, counterion adsorption and overcharging are not well understood. Evidently, there is a significant need to characterize local ion organization near electrified graphene interfaces and elucidate the overcharging process.

Previous studies have experimentally[8-11] and computationally[12,13] considered the graphene-ionic liquid interface in attempts to characterize the EDL relevant to energy storage research.[14] Ionic liquids can adsorb and organize to form an extended EDL, which is undesirable for high charge storage.[10] The transition from overscreening to molecular crowding on the working graphene electrode has also been deemed important.[15] Similar results have been observed for lithium electrolytes.[14]

Fewer works have considered graphene in aqueous environments. Early computational studies reported overcharging or surface-charge amplification for graphene[16] and non-graphene[17] materials alike. These papers consider a complicated EDL structure mechanism. In a simplistic monovalent system, the Poisson-Boltzmann equation describes a typical diffuse layer with exponential decay charge distribution,[17] which has been experimentally observed using x-ray standing waves in a model system.[18] Multivalent ions can induce adsorption of counterions, i.e. the Stern layer, which can create complex layering via charge reversal with respect to the charged surface.[17] These phenomena are being actively considered through computational works.[19, 20] Although the complex layering with aqueous EDLs has been documented for nearly 40 years,[17] little experimental information about the specific mechanism is available for multivalent ions and graphene surfaces.

Surface force apparatus (SFA) experiments boast interfacial-specificity and have provided useful information about EDL structures near mica,[21] which has an intrinsic surface charge and is native to the SFA setup. Experiments using SFA and graphene have been used to determine the physical properties of the interface, including friction and adhesion in a liquid medium.[22] Synchrotron x-ray reflectivity (XR) is suited to probe water and metal ion behavior in the EDL, as high energy XR can access buried interfaces. The measurement at the specular condition yields information about the total electron density profile (EDP) perpendicular to the reflective surface. A low-angle XR study with plain graphene in a buffer solution hinted at ion adsorption.[23] High-resolution XR studies on plain[2, 24] and non-covalently functionalized[25] graphene in water have observed a low-density gap in between the graphene and interfacial water, as well as a distinct layer of adsorbed water that behaved differently compared to the bulk. XR studies of lithiated graphene have revealed information about solid electrolyte interphase formation.[26, 27]

In this paper, we utilize *in situ* high resolution XR, also known as crystal truncation rod (CTR), and resonant anomalous x-ray reflectivity (RAXR) to probe epitaxial graphene-water interface with and without $YCl_3$ to understand water behavior, ion adsorption, and EDL composition in a multivalent charged system. These epitaxial graphene samples are free from transfer substrate residues, which can complicate the EDL revealed with XR.[23] CTR measurements yield the total EDP perpendicular to the graphene surface while RAXR combines high resolution reflectivity with element specificity to provide the EDP specific to adsorbed yttrium ions ($Y^{3+}$)[28-31] (Figure 1). We deploy a custom electrochemical x-ray transmission cell[32] for these measurements that allows us to apply a potential across the graphene surface and probe ion adsorption. The results show $Y^{3+}$ adsorption onto negatively charged graphene, the magnitude of which can be controlled via the applied voltage. We find a small portion of $Y^{3+}$ ions intercalate into the graphene while most of the ions organize into a classical diffuse layer with unexpectedly higher ion coverage. We note these measurements are the first of their kind on a model graphene aqueous system with controlled applied potential and demonstrate the power of interfacial x-ray scattering techniques for determining potential-dependent variations of EDL structure at the graphene-aqueous solution interfaces.

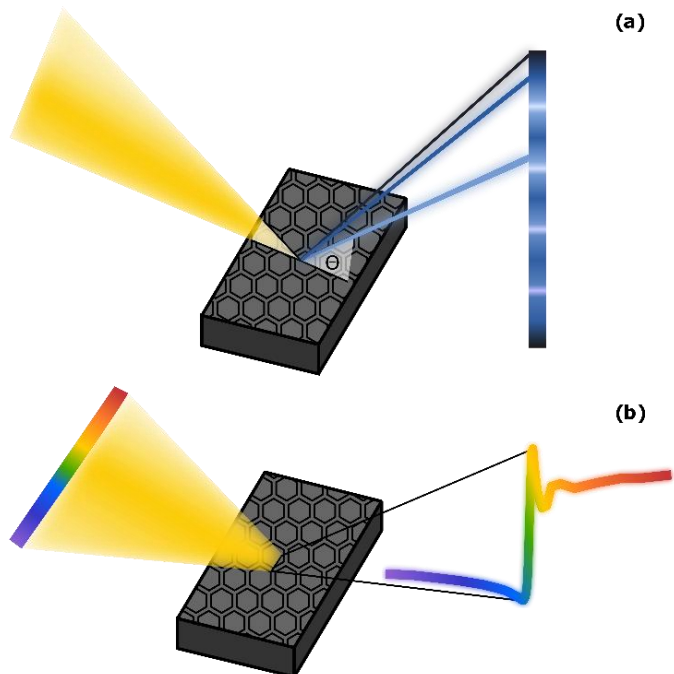

Figure 1. Experimental setup of high-resolution crystal truncation rod (CTR) (a) and resonant anomalous x-ray reflectivity (RAXR) (b) techniques. CTR measures changes in reflectivity (represented as blue-gray color gradient) as a function of momentum transfer ($Q = 4\pi(\sin\theta)/\lambda$, where $\theta$ is the angle between the x-ray beam and the sample surface and $\lambda$ is the x-

ray wavelength) at a fixed photon energy (*E*). RAXR considers changes in reflectivity (represented as rainbow color spectrum) as a function of *E* at a fixed *Q*.

**Materials and Methods**

*Materials*

We prepared solutions using yttrium (III) chloride hexahydrate (Sigma Aldrich, $YCl_3 \cdot 6H_2O$, 99.99% purity) and ultrapure water with a resistivity of 18.2 MΩ•cm (Millipore, Synergy Water Purification System). Chemicals were used as received without additional purification. At the considered concentration, 10 mM, Y is largely $Y^{3+}$ in solution.[33] We purchased epitaxial graphene samples (3 mm x 10 mm) on perpendicularly cut 4H SiC from GraphenSiC as a custom order. Atomic force microscopy data show the expected step-like terrace for 4H SiC when imaged perpendicular to the later x-ray beam path (Supporting Information). Samples contain 2-3 layers of graphene, confirmed with Raman spectroscopy (Supporting Information).

After briefly sonicating in alcohol to remove possible airborne hydrocarbon contaminants, we loaded the graphene sample into a custom electrochemical x-ray transmission cell, described in detail elsewhere.[32] Briefly, two springs connected to two separate pins within Teflon shells to prevent exposure to the surrounding liquid held the graphene sample in place on a plastic platform. These pins applied the desired potential across the graphene surface, i.e., the working electrode. An Ag/AgCl reference electrode (Edaq, Leakless Miniature Ag/AgCl Reference Electrode) was inserted near the graphene. A glassy carbon counter electrode (Alfa Aesar, Glassy Carbon Rod Type 1, 2 mm diameter) was inserted near the graphene as well. These working, reference, and counter electrodes were then connected to a potentiostat (Gamry, Interface 1010E). A typical cyclic voltammetry (CV) measurement for graphene in 10 mM $YCl_3$ is shown in the Supporting Information. The sample shows capacitance in the voltage range of -0.2 – 0.6 V, calculated as 0.202 $C/m^2$. The larger drop in current past -0.2 V likely indicates the onset of water splitting, although hydrogen evolution is not expected to begin until -0.6 V versus Ag/AgCl for pure water.[34] No bubbles were noticed during data collection and the effects of water splitting on our measurements and analysis are likely negligible. The theoretical capacitance of single layer graphene[35] is ~0.1 $C/m^2$ although this value changes for multilayer graphene[36] and assumes an ideal electrolyte with negligible kinetic effects.

*X-ray reflectivity measurements and fitting*

We probed the graphene-liquid interface at the molecular level using high resolution x-ray reflectivity measurements. Experiments were conducted at beamline 33-ID-D at the Advanced Photon Source. The incident beam, in a typical flux of ~$10^{13}$ photons/sec, was focused using a toroidal x-ray mirror to ~0.1 mm (v) x 1 mm (h) at the sample position. Specular *in situ* CTR data were collected by measuring the reflectivity, i.e., the fraction of incident x-ray light reflected from the epitaxial graphene, at a fixed energy (*E* = 20 keV) as a function of momentum transfer $Q = 4\pi (\sin\theta)/\lambda$ where $\theta$ is the incident angle and $\lambda$ is the monochromatic x-ray beam wavelength.[37] RAXR data were collected by measuring the reflectivity at fixed *Q* values over an energy range[38-40] near yttrium's *K* absorption edge,[41] 17.039 keV. To ensure consistency and reproducibility, we repeated measurements in different locations on the same sample. For the voltage-dependent RAXR measurements, we collected reflectivity data using non-sequential voltage steps. Voltages were held for at least 1 minutes before data collection.

*Crystal truncation rod fitting*

We fitted the CTR data with a parameterized structural model that consists of an ideal 4H SiC substrate with 2 relaxed 4H layers at the interface, a buffer layer of graphene grown on top of the substate, 3 layers of graphene with varying coverage, 1-2 layers of adsorbed molecules depending on the sample, i.e. graphene in water versus graphene in 10 mM $YCl_3$, and bulk water, as considered with a distorted layered model[37] (Supporting Information). Details about this fitting methodology are given elsewhere.[37] For graphene in water, we used one layer of adsorbed species. For graphene in 10 mM $YCl_3$, we use two layers of adsorbed species.

*Resonant anomalous x-ray reflectivity fitting*

We fitted the RAXR data using a model-independent method which describes the total structure factor ($F_Y$) of the yttrium ions with[30, 40]

$$F_Y(Q,E) = A_R(Q)exp(iP_R(Q)) \tag{1}$$

where $A_R$ and $P_R$ are its amplitude and phase at specific $Q$. At $Q << 1$, the $A_R$ and $P_R/Q$ can be used to describe the resonant atom occupancy and position, respectively (Supporting Information). From these fits, we determined an appropriate structural model to describe Y adsorption at the graphene interface composed of one layer of adsorbed ions modelled as a Gaussian peak and a diffuse distribution of yttrium beyond the surface, expressed as an exponential decay profile. Details are provided in the Supporting Information.

## Results and Discussion

*Water-Graphene Interface*

We first consider the epitaxial graphene-liquid interface using CTR. Figure 2 shows data collected for graphene in water at 0 V and graphene in 10 mM YCl$_3$ at -0.5 V. CTR signal for graphene in 10 mM YCl$_3$ at additional voltages did not differ significantly, meaning that the interfacial structure accessible via typical non-resonant reflectivity in the measured $Q$ range does not vary with applied potential (Supporting Information). Interfacial changes from ion adsorption as induced with voltage are probed using RAXR, which is sensitive to Y ions and explained in the next section.

Two Bragg peaks are observed at $Q$ = 2.5 and 5 Å$^{-1}$, or $Q = 2\pi L/c_{SiC}$ where $L = 4n_B$, $L$ is the Bragg index in reciprocal lattice units (rlu) for integer values of $n_B$, and $c_{SiC}$ is the lattice spacing parameter measured to be 10.08 Å for 4H SiC[42] (Figure 2a). Graphene peaks appear at $Q$ = 1.75 and 3.50 Å$^{-1}$. Forbidden Bragg peaks are observed solely due to the defects in the SiC substrate[43] (rather than at the interface), and were excluded from the analysis. Noticeably, the CTR signal in the very low $Q$ region differs between the water and 10 mM YCl$_3$ samples, and the first graphene peak at 1.75 Å$^{-1}$ is shifted slightly, as highlighted in the CTR normalized to the SiC substrate (Figure 2b insets). These differences indicate changes in the interfacial structure.

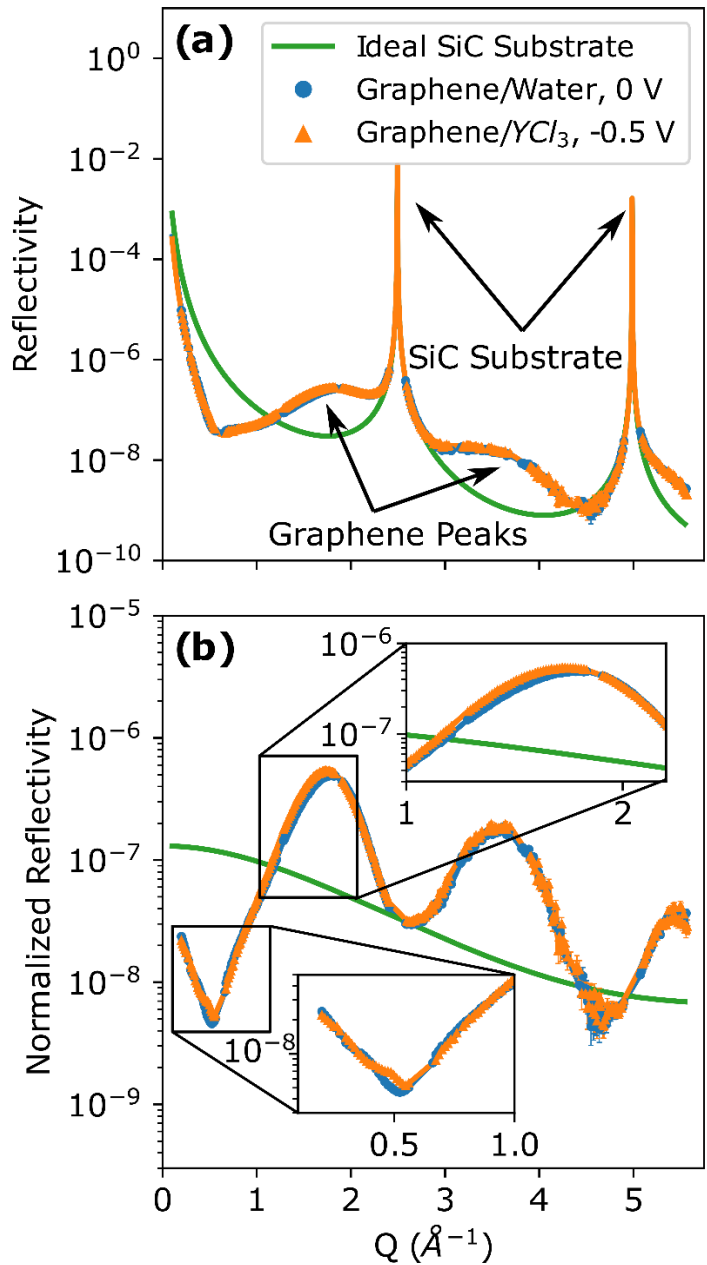

Figure 2. Crystal truncation rod reflectivity (a) and reflectivity normalized to the generic CTR shape ($= 1/[Q \sin(Qd/8)]^2$) of the SiC substrate (b) over momentum transfer $Q$ for 3 layers of epitaxial graphene on SiC in water (blue) and 10 mM $YCl_3$ (orange) with those calculated for an ideal SiC substrate in air (green). Data are fitted to a structural model, as described in the text. Error bars are derived from experimental counting statistics.[37]

To quantify the changes in the interfacial structure, we fit the CTR data to a structural model[24, 25] (Equation S1) considering an ideal 4H SiC substrate with 2 relaxed 4H layers at the interface, a buffer layer of graphene grown on top of the substrate ($G_0$), 3 layers of graphene with varying coverage ($G_1$, $G_2$, and $G_3$), adsorbed molecules to represent adsorbed water species, and bulk water, as considered with a distorted layered model. Essentially, each of these subset structures represents a series of specific "slabs" with known thicknesses, roughness values, and electron densities with additional fine details in the atomic components within the layer, which are distinguished by the spatial resolution of the data (i.e., ~0.6 Å = $\pi/Q_{max}$, where $Q_{max}$ is the maximum $Q$ value per each dataset). By building a reasonable structural model, we can determine occupation factors ($O$) for graphene and adsorbed ion/molecule layers, as explained in the Methods section. We use one adsorbed layer for graphene in water and two adsorbed layers for graphene in 10 mM $YCl_3$ to represent increased complexities owing to adsorbed yttrium. Table S1 shows the relevant fitting parameters derived from the best-fit models. We also considered other model fits, such as fewer layers of graphene, no adsorbed species, etc., in the Supporting Information.

From the total structure factor, we calculate an electron density profile for both graphene in water and graphene in 10 mM $YCl_3$ (Figure 3). Both samples have nearly identical SiC substrates and $G_0$ layers, as expected for structures below the exposed interface. The first two graphene layers, $G_1$ and $G_2$ have 100% and 42% coverage, respectively, in both cases. The topmost graphene layer $G_3$ has smaller partial coverage, i.e., 9% for graphene in water and 4% for graphene in 10 mM $YCl_3$. These differences in graphene layer coverage are typical for multilayer epitaxial graphene samples where it is difficult to control the absolute number of layers across the sample during growth.[44]

Taken together, these partial coverage values mean that there are different layers across the graphene surface that are exposed to water and possibly yttrium ions. In the case of graphene in water, water can adsorb onto $G_1$, $G_2$, and $G_3$. This adsorption uniquely differs from bulk water behavior, as it is more structured and generates observable changes in reflectivity.[37] The buffer layer $G_0$ appears at 2.3 Å (Figure 3). Graphene layers appear at 5.7, 9.1, and 12.5 Å (Figure 3), with a consistent reasonable interlayer spacing[45] of ~3.4 Å. The first adsorbed species appears at 8.5 Å from the surface, or 2.8 Å away from $G_1$, and is likely water adsorbed on $G_1$. The height of this water adsorbed on a graphene layer agrees with previously published works.[9, 24] Bulk water approaches $G_1$ at 10.7 Å and is considered using a distorted layer model. Adsorbed species also appear at 11.9 and 15.3 Å and are interpreted as water adsorbed on subsequent partial graphene layers $G_2$ and $G_3$, respectively. The bulk water shows an additional peak at 14.1 Å, corresponding with bulk water contact with $G_2$. Bulk water that is largely featureless dominates the electron density profile after ~20 Å.

The electron density profile for graphene in 10 mM $YCl_3$ shows significant changes caused by the ion adsorption. Two adsorbed species are considered for each graphene layer that is exposed to the solution. Again, water and ions can adsorb onto $G_1$, $G_2$, and $G_3$ because the uppermost graphene layers only have partial coverage. An adsorbed species appears at 8.1 Å, 2.3 Å away from $G_1$, and overlaps slightly with $G_2$. A second peak shows at 10.7 Å, 4.9 Å away from $G_1$. These peaks may represent water and/or yttrium ions that have adsorbed onto $G_1$ based on the assumption that only water and yttrium will adsorb to the sample under -0.5 V applied potential. Our data may be less sensitive to chloride ions, which are likely repelled from the negatively charged surface although we cannot fully exclude the possibility of their adsorption by positional correlation with interfacial Y ions, e.g., to compensate the potential overcharging by adsorption of the multivalent cations.[33, 46] Given the similar position for the graphene in water case discussed previously, it is likely that the adsorbed species at 8.1 Å is water. However, the different adsorption heights (2.8 vs. 2.3 Å) as well as distribution widths (0.2 Å vs. 0.4 Å) may be affected by the applied voltage as our data for graphene in water were measured at 0 V and our data for graphene in 10 mM $YCl_3$ were measured at -0.5 V. We suspect that water alignment differs with applied voltage, but its orientation, especially the location of hydrogen atoms, cannot be fully resolved in the CTR data.

We find two adsorbed peaks for $G_2$ at 11.8 and 14.1 Å, and two adsorbed peaks for $G_3$ at 15.2 and 17.5 Å. The interpeak spaces and distance from each graphene layer are held constant within the model. Bulk water approaches the sample at 12.8 Å and covers $G_3$ – consistent with a small partial $G_3$ coverage that allows water to approach the layer. The bulk water model shows more distinct layering, as evidenced by the prominent electron density peaks at 12.8, 16.2, and 20 Å, compared to the graphene in water without applied potential. This extended layering implies that the additional adsorbed species also affect water behavior over an extended range in height at the electrified graphene in the $YCl_3$ solution.

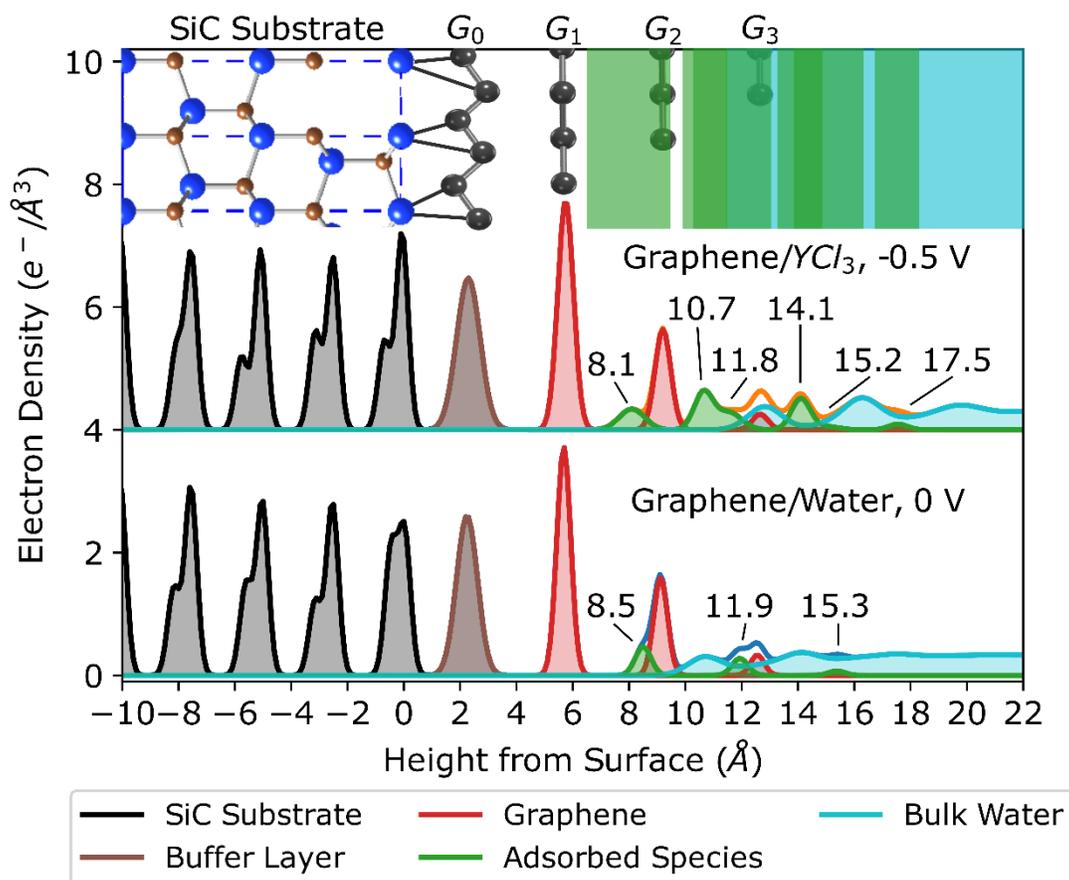

Figure 3. Electron density profile as a function of height away from the sample surface derived from modelled total non-resonant structure factor for graphene in water (blue, bottom) and graphene in 10 mM $YCl_3$ (orange, top). Model considers 2 layers of 4H SiC substrate (black), a buffer layer of epitaxial graphene grown on the substrate ($G_0$, brown), 3 layers of graphene with partial coverage ($G_1 – G_3$, red), adsorbed species (green), and bulk water (cyan). The graphene data in 10 mM $YCl_3$ are offset by +4 $e^-/Å^3$ for clarity.

Many of the small electron-density features overlap with each other in the laterally averaged representation. To visually separate the distribution of individual adsorbed species from one another, we present the electron density profile of the solution species only on a single layer of graphene with full coverage ($G_s$), as calculated using the fitted parameters from our experimentally measured data with multilayer graphene (Figure 4). In the case of graphene in water, we clearly see an adsorbed species 2.8 Å from the graphene layer, which we attribute to adsorbed water molecules. Around 5.2 Å, we observe a layer of ordered water, which is modelled as the first water peak of the distorted layer model, above which the water is largely featureless. For graphene in 10 mM $YCl_3$, two adsorbed species appear 2.3 and 4.9 Å from the graphene, which are similar in height to those in water. Therefore, we posit that these species are mainly adsorbed water whereas small differences in their distribution reflect the changes in the surface hydration structure, owing to the (increased) negative charge of the graphene surface under applied potential and likely the presence of adsorbed cations to compensate the charge. The solution structure also shows more distinct layering, as evidence by the electron density peaks at 10.8 and 14.5 Å and the electron density dips near 8.8 and 12.5 Å. These extended structures have also been observed in $YCl_3$/muscovite mica systems.[33] The (001) surface of the mica has a net negative charge, similar to the electrified graphene surface in our study, with a charge density of ~-0.32 C/m$^2$ that is larger than the value for graphene that we measured by cyclic voltammetry (~-0.2 C/m$^2$).

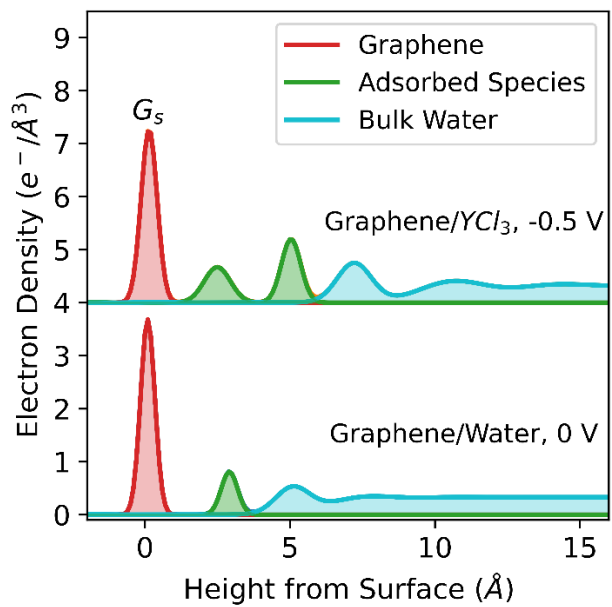

Figure 4. Electron density profile as a function of height from the graphene surface calculated for a single layer of graphene with total coverage ($G_s$) in water (blue, bottom) and in 10 mM YCl$_3$ (orange, top) using the fitted adsorbed species and water parameters of a modelled total non-resonant structure factor. The model considers an adsorbed species (green) and bulk water (cyan) among other parameters not shown here. The graphene data in 10 mM YCl$_3$ are offset by +4 $e^-/Å^3$ for clarity.

*Element-sensitive studies of Y$^{3+}$ adsorption*

To better understand the distribution of yttrium near the graphene surface, we perform RAXR measurements on graphene in 10 mM YCl$_3$ at -0.5 V (Figure 5). RAXR is elementally specific and measures changes in reflectivity signal from a target resonant metal by tuning the x-ray energy around the x-ray absorption edge of the element. We first utilized a model-independent approach where the $Q$-dependent variations in RAXR signal (Figure 5) are explained using the amplitude, $A_R$ and phase, $P_R$ of the partial structure factor of Y adsorbed at the interface. Changes in these values over $Q$ produce different spectral heights and shapes, as explained in detail elsewhere.[40] Briefly, at $Q \ll 1$, the fitted value $A_R$ describes the occupancy of the resonant atoms and fitted value $P_R$ divided by $Q$ describes the average position of the resonant atom relative to the reflective surface. We fitted our RAXR data using these parameters (Equation 1) via the least sum of squares.

Figure 5 shows RAXR signals observed at $Q$ from 0.19 to 0.85 Å$^{-1}$, indicating sorption of yttrium near graphene at -0.5 V. The spectral shape gradually changes with increasing $Q$, providing quantitative information on the distribution of the resonant atom at the interface. At $Q$ = 0.19 Å$^{-1}$, the lowest $Q$ investigated in this study, $A_R$ is 1 Y$^{3+}$ per 55 Å$^2$, which provides the estimation of the minimum Y coverage. That is, the actual coverage is presumably higher than this estimation but is difficult to determine precisely because of the difficulty of measuring RAXR from our small sample at lower $Q$.

$Q$-dependent variation in $P_R/Q$ provides insight into the average height and shape of the overall Y profile. The $P_R/Q$ at the lowest $Q$ is ~28 Å, corresponding to the average height of sorbed Y measured from the top surface of SiC. This estimated height is substantially higher than the height expected for adsorbed ions by simple inner-sphere surface complexation even when the height shifts by the presence of the graphene layers are considered. The $P_R/Q$ gradually decreases with increasing $Q$, indicating asymmetry of the overall profile. The negative gradient indicates that the profile has a positive skew in which its high-density peak is located nearer the surface while its diffuse tail is extended farther from the surface. At $Q \geq 0.69$ Å$^{-1}$, the $P_R/Q$ value approaches ~10 Å, which is close to the average height of the graphene layers at the surface.

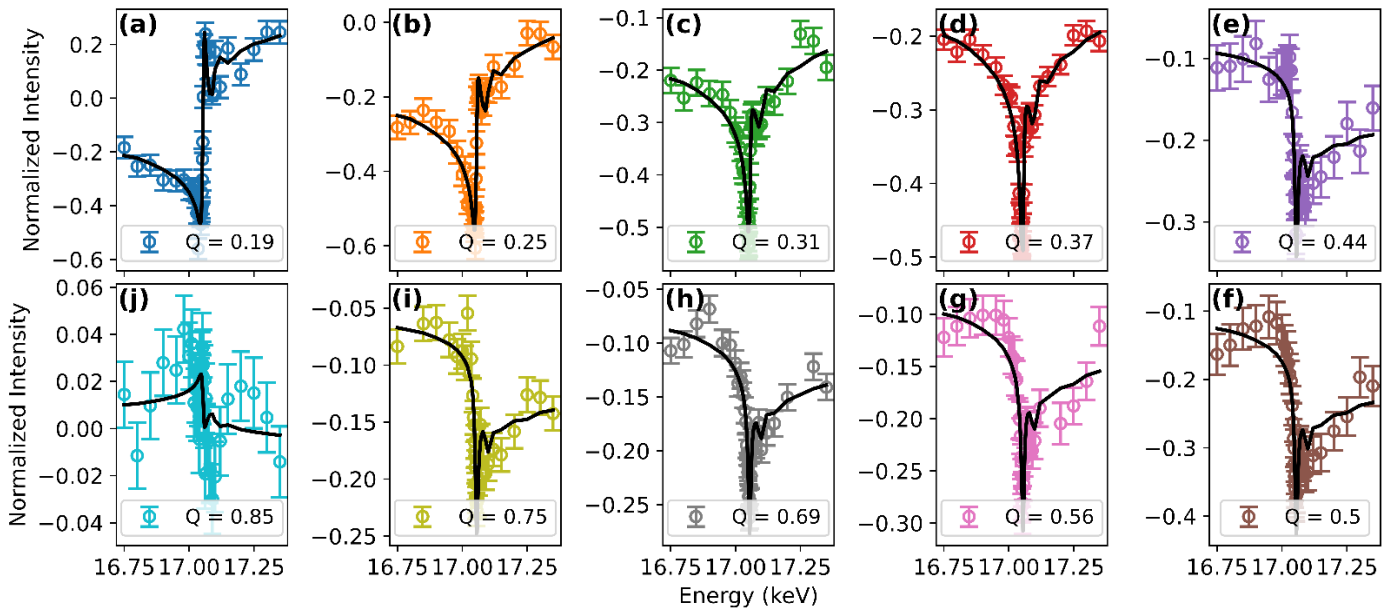

Figure 5. Normalized resonant anomalous x-ray reflectivity (($|F_{tot}(Q,E)|^2 - |F_{NR}(Q)|^2)/(2|F_{NR}(Q)|)$)[47, 48] and fits (black lines) of graphene in 10 mM YCl$_3$ under an applied potential of -0.5 V as a function of incident beam energy, varied around the $K$-edge of solution yttrium, 17.039 keV, at different momentum transfer $Q$ (Å$^{-1}$) values (panels). Data are fitted via a model-independent analysis where collected signal is described using a non-resonant structure factor known from crystal truncation rod analyses, an anomalous dispersion term determined from x-ray absorption measurements, and fitted amplitude and phase to describe the magnitude and shape of the spectrum. Error bars correspond to the 1-sigma uncertainties of the data determined via counting statistics.

The information obtained from the model independent analysis can be further quantified with a structural model that includes a layer of adsorbed yttrium close to the graphene surface and a diffuse profile of yttrium farther from the surface. A Debye length of 12.4 Å calculated for 10 mM YCl$_3$ was used for our model.[49] This value is calculated independently from the x-ray measurements and its effects on the fit results have been investigated by fitting the data with different Debye lengths (Supporting Information). The fitted amplitude and phase values obtained from the model-independent analysis and the expected values, calculated from Equations 1 and S5, determined from this structural model are plotted over $Q$ in Figure 6. Via the structural model fit, an adsorbed layer located at 1.00 ± 0.09 Å with an rms width of 2.2 ± 0.1 Å accounts for a smaller fraction of Y sorption, with an occupancy of 1 Y$^{3+}$ per 125 ± 7.6 Å$^2$. This ion distribution is below the height of the first graphene layer ($G_1$) and may represent a fraction of Y sorbed through the defects of the graphene layers. We considered other fits without this buried layer and were unable to obtain satisfactory $\chi^2$ values. A diffuse layer whose peak density is located 19.2 ± 0.2 Å from the SiC surface and tail extends to ~50 Å has a significantly larger occupancy of 1 Y$^{3+}$ per 11.4 ± 1.6 Å$^2$. The front of this profile, i.e. the side closest to the surface, begins around ~9 Å, which is ~3 Å apart from the $G_1$ layer and similar to the height of the $G_2$ layer. The peak electron density of adsorbed yttrium is located at 20 Å, which is more than 5 Å farther from the top graphene layer. This large spacing can be explained by adsorption of yttrium ions as outer-sphere surface complexes that maintain their hydration shells intact. Dominant outer-sphere adsorption has been observed for ions having large hydration free energies.[33, 50] For example, the Gibbs free energy of hydration for $Y^{3+}$ is circa -3500 kJ/mol, significantly larger than those of mono- and di-valent cations in similar ionic radii (e.g., -365 and -1505 kJ/mol for $Na^+$ and $Ca^{2+}$, respectively[51]). In contrast, only a small fraction adsorbs closer to the surface, indicating minimal contribution from inner-sphere surface complexes. This limited inner-sphere adsorption can be attributed to the absence of any functional groups (e.g., terminal ligands that are prevalent in many solid surfaces including common oxides) on the chemically inert graphene surface to facilitate dehydration. While the negative surface charge induced by the applied potential electrostatically attracts $Y^{3+}$ ions to the surface, it may not be strong enough to break the ion hydration shell.

The experimentally determined coverage of $Y^{3+}$ ions is significantly higher than the amount needed to compensate the surface charge induced by the applied potential, estimated as 1 $Y^{3+}$ per 236 Å$^2$ via CV measurements. It is known that multivalent ions can show overcharging due to ion-ion correlations at charged interfaces.[33, 52, 53] The apparent overcharging is likely compensated by $Cl^-$ ions co-adsorbing in the diffuse layer to offset the excessive cation charge. The overcharging we observe is unexpected by classical EDL theories and highlights the necessity of thorough characterization.

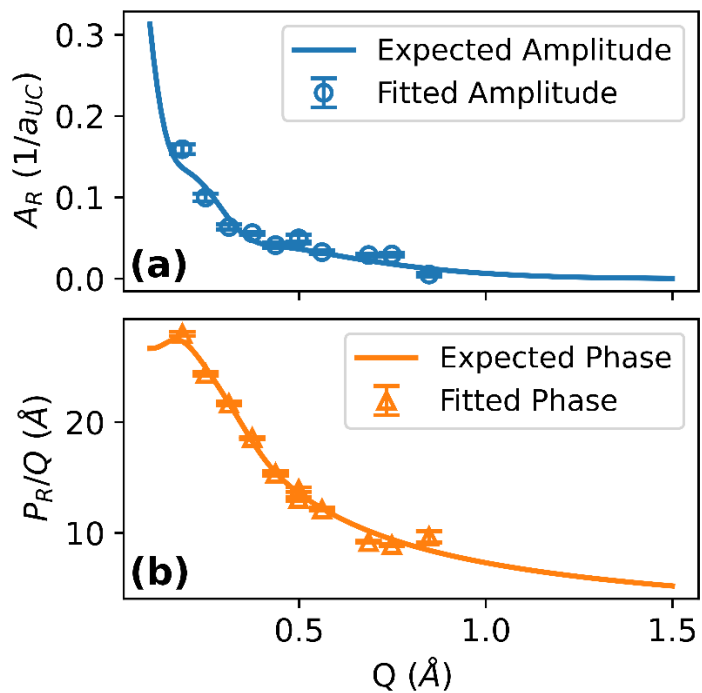

Figure 6. Fitted partial structure factor amplitudes ($A_R$, a) and phases ($P_R$, b) plotted over $Q$ obtained via model dependent fits of resonant anomalous x-ray reflectivity of graphene in 10 mM $YCl_3$ under an applied potential of -0.5 V. Expected amplitude and phase values (lines) are calculated from a structural model considering an adsorbed layer of ions close to the surface and a diffuse profile of ions farther from the surface. Error bars correspond to the 1σ uncertainties derived from the analysis. $A_R$ are provided with respect to the area of the unit cell ($a_{UC}$).

To highlight the ion coverage within the observed diffuse layer, we plot the electron density profile obtained from our model-dependent RAXR fitting results alongside the electron density profile obtained from our CTR fit for graphene in 10 mM $YCl_3$ in Figure 7. Again, a small portion of the total electron density comes from $Y^{3+}$ ions adsorbed on the SiC crystal below the graphene layers exposed to the bulk solution. Most of the adsorbed $Y^{3+}$ ions exist within the extended layer that continues into the bulk (Figure 7, inset). These data show an unexpectedly high ion coverage. While our data allow robust determination of the overall shape and the location of the peak density of the Y profile, they have limited sensitivity to its extension (i.e., Debye length) because of the absence of the data at lower $Q$. Instead, we emphasize that most yttrium ions are located within ~3 nm of the solution near the electrified graphene surface.

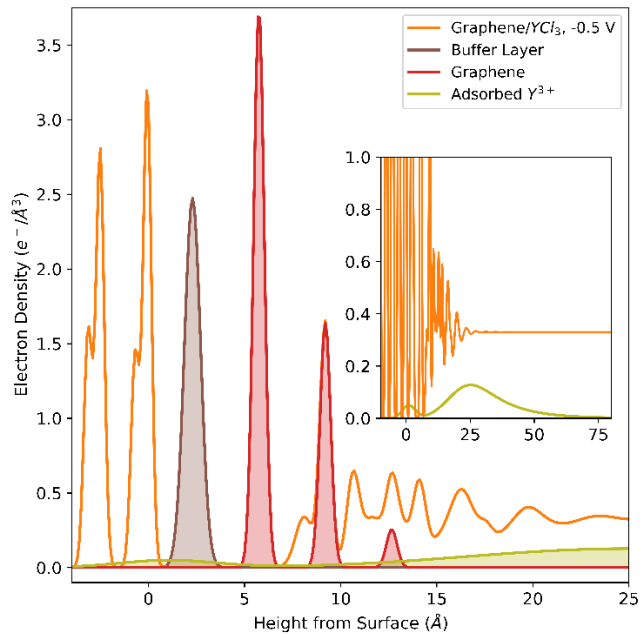

Figure 7. Total electron density profile derived from CTR for graphene in 10 mM YCl$_3$ (orange) plotted with the Y electron density profile derived from RAXR (yellow) as a function of height from the SiC surface. Electron density profiles for the extended interfacial region are provided in the inset, which has the same axis labels and units.

*Potential-dependent Y$^{3+}$ adsorption*

We probed trivalent adsorption as a function of applied potential by collecting RAXR data for graphene in 10 mM YCl$_3$ at different voltages and at a fixed momentum transfer of $Q$ = 0.5 Å$^{-1}$ (Figure 8). These data were analyzed using the model-independent method. Minimal magnitude in normalized RAXR signal appears at small negative voltages, V = -0.10 to -0.40 V, indicating nearly no sorbed yttrium. The signal rapidly increases at V = -0.44 to -0.60 V, as shown by the distinct magnitude change. Because the RAXR signal is only produced by yttrium at the interface, Y$^{3+}$ must accumulate near the surface as the graphene becomes more negatively charged. These changes are quantified using the amplitude and phase values plotted over the applied voltage in Figure 9. The very low amplitude values at small negative voltages indicate that the resonant ion has low occupancy. In V = -0.1 to -0.4 V, the $P_R/Q$ data show an apparent decrease with increasing the magnitude of applied potential. This trend normally indicates a decrease of the average height of adsorbed Y$^{3+}$. However, its statistical significance is difficult to determine because of the large uncertainties of the data. These large uncertainties for the phase values are due to the difficulty in determining the shape of the spectra with no (or little) signals. Gradually increasing the magnitude of the negative potential has little impact on the amplitude until V = -0.40 V where a statistically non-zero amplitude was determined. The phase values remain constant, within the error bars, for the remainder of the tested voltages, meaning that the adsorbed ions maintain their average position (and the overall shape of the profile by inference). The derived amplitude slowly increases as the applied voltage becomes more negative, implying that more yttrium ions adsorb to the graphene. We confirmed the reproducibility by collecting multiple datasets and varying both $Q$ and the applied voltage non-sequentially. Signal intensity also did not change over time, implying the system was stable in the x-ray beam.

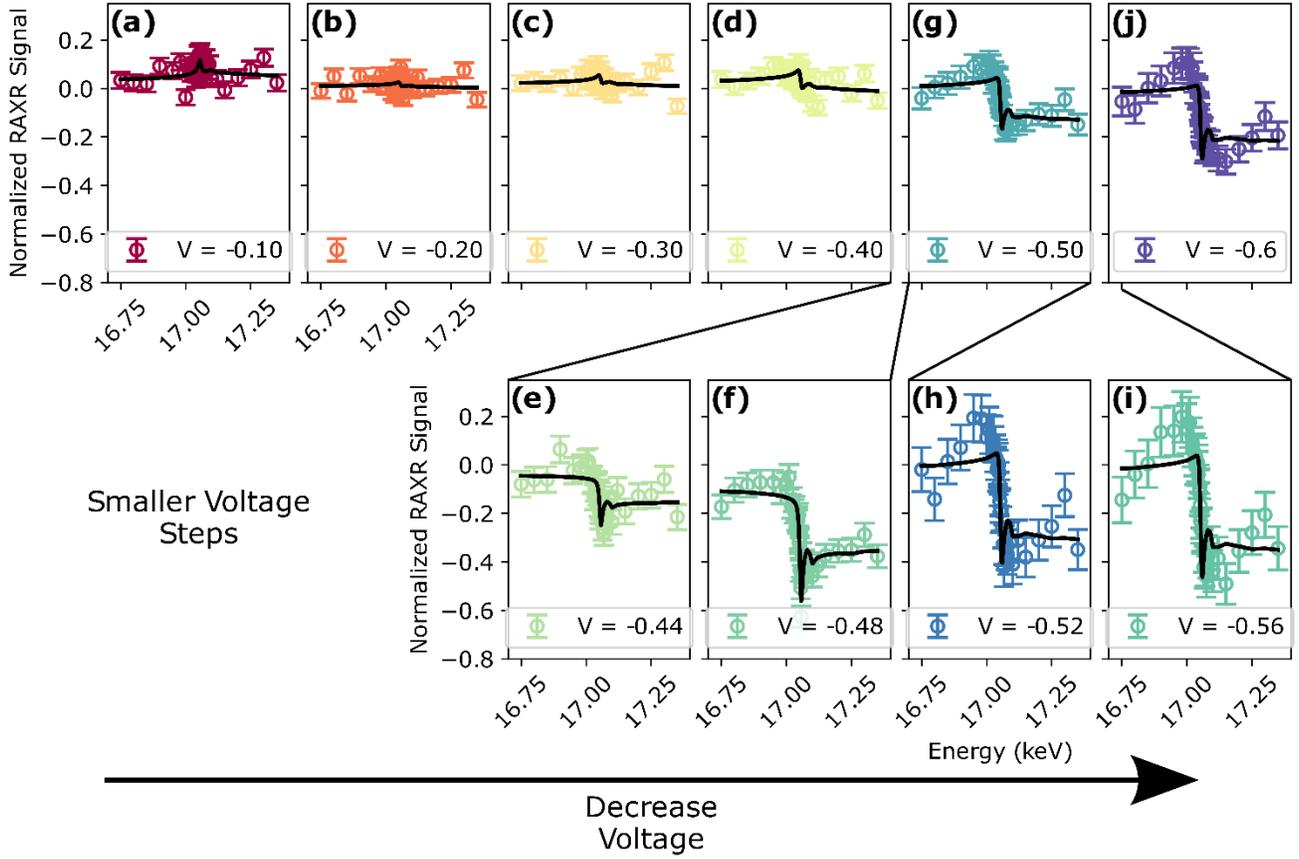

Figure 8. Resonant anomalous x-ray reflectivity and fits (black lines) of graphene in 10 mM YCl₃ under varied applied potentials (A-J) as a function of incident beam energy, varied around the *K*-edge of solution yttrium, 17.039 keV. Data are fitted via a model-independent analysis where collected signal is described using a non-resonant structure factor known from crystal truncation rod measurements, an anomalous dispersion term known from x-ray absorption measurements, a fitted amplitude to describe the occupancy of the resonant species, and a fitted phase to describe the position of the resonant species. Data are collected at fixed momentum transfer value $Q$ = 0.5 Å$^{-1}$. Error bars are determined via counting statistics.

To express ion adsorption phenomena as a function of the applied potential (*V*), we fitted via the least sum of squares the obtained amplitude values (*A*) to a pseudo-Langmuir isotherm

$$A(E) = A_0 + (A_m - A_0)\frac{e^{(E_{1/2}-E)/kT}}{1 + e^{(E_{1/2}-E)/kT}} \quad (2)$$

where $A_0$ and $A_m$ are the minimum and maximum observed amplitudes, respectively, $E_{1/2}$ is the energy at which $A(E) = (A_m - A_0)/2$, $k$ is the Boltzmann constant, and *T* is temperature (298.15 K). This equation assumes that the potential-dependent variation in adsorbed ion coverage is controlled by an apparent free energy of adsorption at the surface.[33] This fitting yields $E_{1/2}$ = -0.46 ± 0.01 eV. A baseline coverage on the graphene surface is required to facilitate this adsorption and some ions appear to adsorb at potentials as low as -0.1 V via our fitted *P/Q* plot. The obtained amplitude is not substantial until V = -0.4 V where it begins to increase. This amplitude approaches a plateau around V = -0.5 V where magnitude of the amplitude approaches A = 0.06.

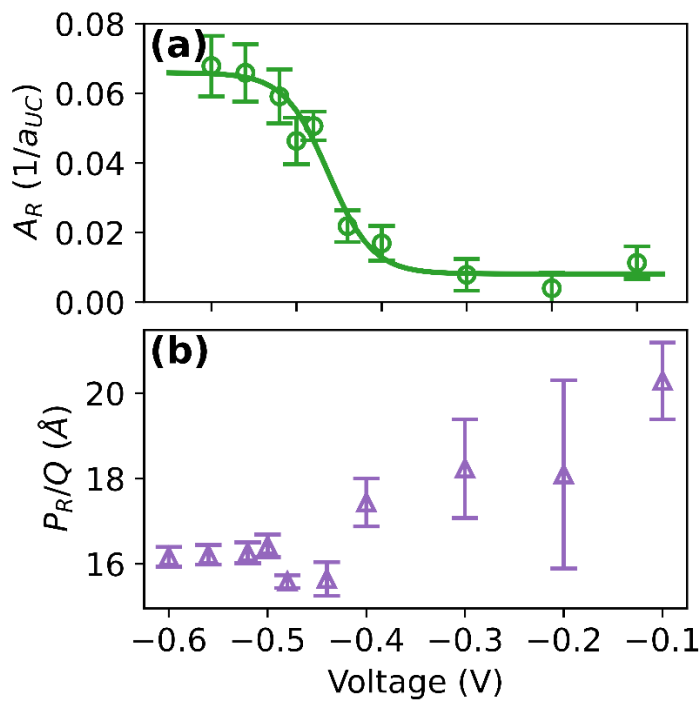

Figure 9. Fitted partial structure factor amplitudes ($A_R$, a) and phases ($P_R$, b) plotted over varied applied voltage obtained via model independent fits of resonant anomalous x-ray reflectivity of graphene in 10 mM $YCl_3$ measured at $Q$ = 0.5 Å$^{-1}$. Error bars correspond to the 1σ uncertainties derived from the analysis. The partial structure factor amplitude was fitted to a pseudo-Langmuir isotherm, provided in Equation 2. $A_R$ are provided with respect to the area of the unit cell ($a_{UC}$).

## Conclusions

The structure of the EDL in aqueous systems with charged graphene is important for both basic science and applied graphene technology endeavors alike. Adsorption of multivalent ions in particular can overcharge graphene. Despite the importance of overcharging in system efficiency, relatively little is known about the mechanism. We present CTR and RAXR data studying the solid-liquid interface of electrified graphene in dilute $YCl_3$ and show $Y^{3+}$ adsorption as a function of applied potential. Our x-ray data reveal both yttrium ion coverage and proximity to graphene and demonstrate $Y^{3+}$ absorption at a static negative potential. From the model analysis, we find most absorbed $Y^{3+}$ exist within 2 nm of the charged graphene surface. Surprisingly, the occupancy of $Y^{3+}$ ions is very high and is well beyond the charge required to neutralize the graphene surface. This diffuse layer has a typical shape but unexpectedly large cation charge.

When we vary the applied voltage, we observe a consistent average $Y^{3+}$ position with respect to the surface and increasing Y coverage with decreasing potential. Taken together, these data support overcharging of the graphene surface. This overcharged diffuse layer we observe can exist if chloride anions are effectively shielding or screening the adsorbed $Y^{3+}$ charge. Our characterization of this EDL provides important molecular information about a relevant aqueous system and paves a way for more detailed studies on overcharged graphene systems.

## Acknowledgements

We thank Paul Fenter for valuable discussions, Tim Fister for the electrochemical cell design, and Zhan Zhang for the help with the synchrotron experiments. The work presented here was supported by the U.S. Department of Energy, Office of Basic Energy Science, Division of Chemical Sciences, Geosciences, and Biosciences, Early Career Research Program to A.J.C. and A.U. and Geosciences Program to S.S.L. under contract DE-AC02-06CH11357 to UChicago Argonne, LLC as operator of Argonne National Laboratory. The use of Sector 33-ID-D and the Advanced Photon Source, an Office of Science User Facility operated for the U.S. Department of Energy (DOE), Office of Science by Argonne National Laboratory, was supported by the U.S. DOE under Contract no. DE-AC02-06CH11357. Use of the Center for Nanoscale Materials, an Office of Science user facility, was supported by the U.S. Department of Energy, Office of Science, Office of

**Supporting Information for**

# Trivalent ion overcharging on electrified graphene

Amanda J. Carr, Sang Soo Lee, Ahmet Uysal*

Chemical Sciences and Engineering Division, Argonne National Laboratory, Lemont, IL 60439, United States

*Corresponding Author Email: auysal@anl.gov

*Epitaxial graphene characterization*

Raman characterization (Figure S1) on the epitaxial graphene samples before x-ray studies revealed the characteristic[1] *G* peak at 1580 cm$^{-1}$ and *2D* peak at 2660 cm$^{-1}$. The *2D*:*G* peak intensity ratio = 0.884, consistent with 2-3 layers of graphene on the sample surface.[2] The lack of the *D* peak around ~1350 cm$^{-1}$ indicates that high quality graphene is present. Atomic force microscopy data (Figure S2) shows the expected step features[3] for 4H SiC.

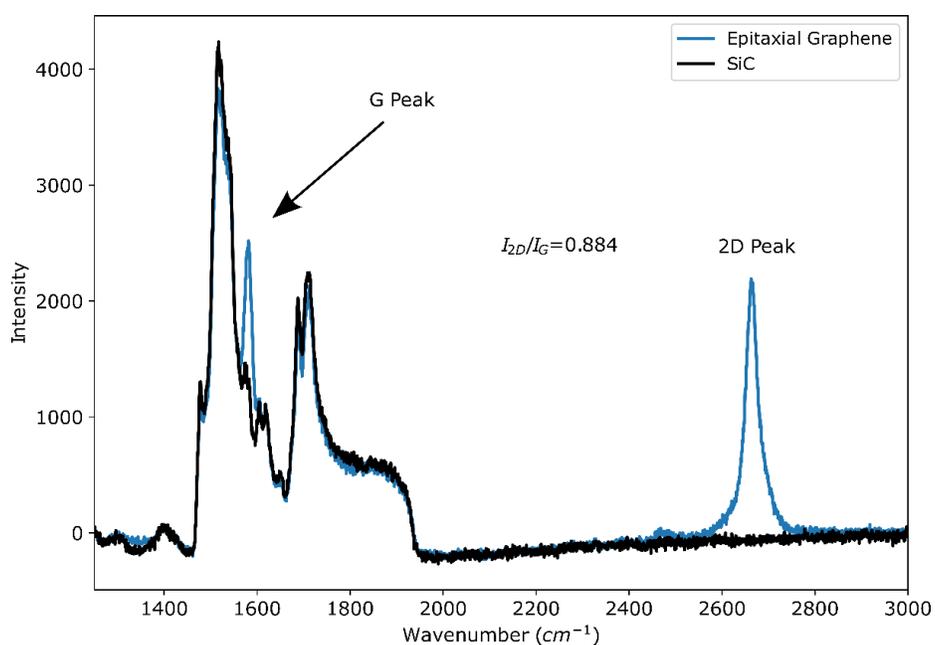

Figure S1. Raman intensity of epitaxial graphene on 4H SiC substrate over wavenumber collected using a 633 nm laser.

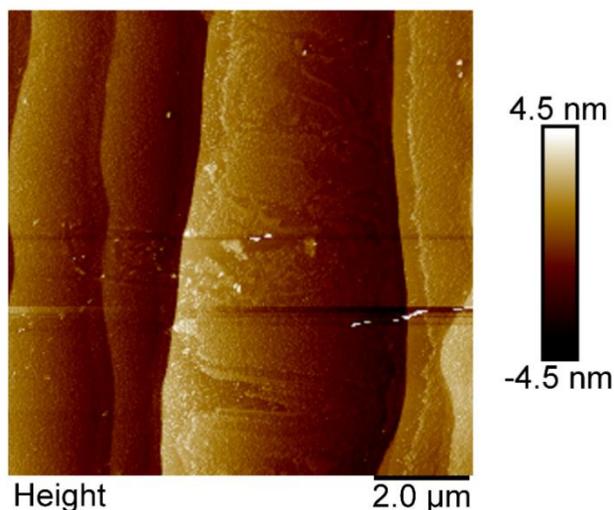

Figure S2. Atomic force microscopy of epitaxial graphene on 4H SiC substrate collected using a Veeco MultiMode 8 SPM.

*Cyclic voltammetry capacitance calculation*

$$C = \frac{\int_{V_f}^{V_I} I \, dV}{2\Delta V \frac{dV}{dt}} \quad (S1)$$

We calculate the capacitance, *C*, of our graphene in 10 mM YCl$_3$ sample using Equation S1 where *I* is the measured current over voltage range $V_I$ to $V_f$ using a step size $\frac{dV}{dt}$. Our data was collected at 5 mV/s and measured from V = -0.6 – 0.6 V. A typical cyclic voltammetry plot is shown in Figure S3. As discussed in the Methods, the sample is largely capacitive and water splitting has a negligible effect on ion adsorption and our analysis.

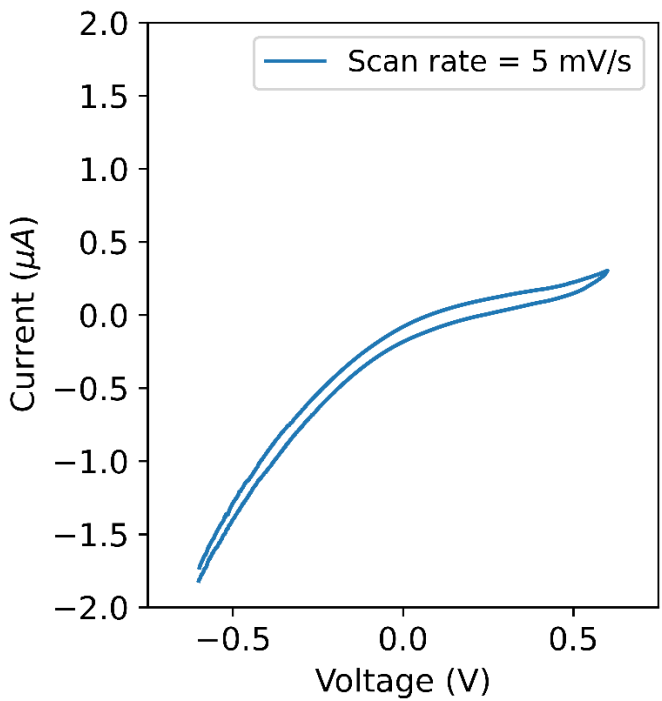

Figure S3. Cyclic voltammetry of graphene in 10 mM YCl$_3$ showing measured current over voltage with respect to an Ag/AgCl reference electrode. Data were collected using a 5 mV/s scan rate and a glassy carbon counter electrode.

*Crystal truncation rod fitting*

Our model calculates the total reflectivity *(R)* via

$$R(Q) = \left(\frac{4\pi r_e}{QA_{UC}}\right)^2 |F_{UC}F_{CTR} + F_{Int} + F_W|^2 \quad (S2)$$

where $r_e$ is the classical electron radius, 2.818 x 10$^{-5}$ Å, and $A_{UC}$ is the area of the unit cell of SiC, 8.22 Å$^2$. The structure factor of the substrate SiC crystal is calculated as a product of $F_{UC}$, the structure factor for the atoms in a unit cell, and $F_{CTR}$, the CTR form factor summed over a semi-infinite crystal, $1/(1 - exp(-iQc_{SiC}))$ for 4H SiC where $c_{SiC}$ is the d-spacing of the (001) plane. The structure factors of the interface and bulk water ($F_{Int}$ and $F_W$, respectively) are calculated over *j* atoms using

$$F = \sum_j O_j f_j(Q) \, exp(iQz_j) \, exp\left(-\frac{(Qu_j)^2}{2}\right). \tag{S3}$$

Here, $f_j(Q)$ is the atomic scattering factor, $O_j$ is the atom occupancy, $z_j$ is the atom height from the crystallographic position of the top Si, and $u_j$ is the atom's rms distribution width. For simplicity, the adsorbed species were modelled as oxygen. We fitted data via the least sum of squares and report $\chi^2$ as a fit quality parameter. Care was taken to ensure limited variable covariance, i.e. Pearson correlation coefficients[4] r < 0.9.

**Table 1**. Crystal truncation rod model fitting results for 3 layers of graphene with variable coverage on a relaxed SiC substrate with adsorbed species and bulk water.

| Fitted Model Parameters | Graphene in Water | Graphene in 10 mM YCl$_3$ |
|---|---|---|
| Graphene layer occupancy (C/A$_{UC}$) | $G_0$: 3.332 ± 0.077 | $G_0$: 3.560 ± 0.033 |
| | $G_1$: 3.153 ± 0.036 | $G_1$: 3.577 ± 0.046 |
| | $G_2$: 1.350 ± 0.00030 | $G_2$: 1.564 ± 0.067 |
| | $G_3$: 0.281 ± 0.018 | $G_3$: 0.244 ± 0.053 |
| Graphene layer rms width (Å) | $G_0$: 0.316 ± 0.0071 | $G_0$: 0.369 ± 0.014 |
| | $G_{n>0}$: 0.147 ± 0.0095 | $G_{n>0}$: 0.196 ± 0.021 |
| Graphene $d$ spacing (Å) | 3.437 ± 0.0097 | 3.455 ± 0.0060 |
| Topmost SiC position (Å) | Si: 1.027 ± 0.035 | Si: 0.938 ± 0.020 |
| | C: 1.173 ± 0.012 | C: 0.935 ± 0.048 |
| Topmost SiC occupancy (atom/A$_{UC}$) | Si: 0.726 ± 0.045 | Si: 1.013 ± 0.052 |
| | C: 1.523 ± 0.079 | C: 1.064 ± 0.14 |
| Adsorbed species 1 position (Å) | 2.817 ± 0.057 | 2.351 ± 0.063 |
| Adsorbed species 1 occupancy (O/A$_{UC}$) | 0.592 ± 0.082 | 0.787 ± 0.045 |
| Adsorbed species 1 peak rms width (Å) | 0.2 (resolution limit) | 0.406 ± 0.074 |
| Adsorbed species 2 position (Å) | -- | 4.900 ± 0.042 |
| Adsorbed species 2 occupancy (O/A$_{UC}$) | -- | 1.042 ± 0.034 |
| Adsorbed species 2 peaks rms width (Å) | -- | 0.269 ± 0.061 |
| Bulk water height (Å) | 4.990 ± 0.19 | 7.073 ± 0.033 |
| Bulk water peak rms width (Å) | 0.556 ± 0.20 | 0.557 ± 0.048 |
| Distance between layered water peaks (Å) | 2.209 ± 0.84 | 3.371 ± 0.074 |
| $\chi^2$ | 0.582 | 0.772 |

**Table S2.** Crystal truncation rod model fitting results for other considered models

| Model Description | Fitted Model Parameters | Graphene in Water | Graphene in 10 mM YCl$_3$ |
|---|---|---|---|
| 3 layers of graphene with partial coverage, bulk water | Graphene layer occupancy (C/A$_{UC}$) | $G_0$: 2.856 ± 0.013 | $G_0$: 2.903 ± 0.063 |
| | | $G_1$: 2.856 ± 0.016 | $G_1$: 3.251 ± 0.061 |
| | | $G_2$: 1.365 ± 0.0098 | $G_2$: 1.808 ± 0.048 |
| | | $G_3$: 0.409 ± 0.0080 | $G_3$: 0.644 ± 0.049 |
| | Graphene layer rms width (Å) | $G_0$: 0.342 ± 0.010 | $G_0$: 0.598 ± 0.056 |
| | | $G_{n>0}$: 0.174 ± 0.0063 | $G_{n>0}$: 0.178 ± 0.0067 |
| | Graphene $d$ spacing (Å) | 3.461 ± 0.0027 | 3.405 ± 0.0061 |
| | Topmost SiC position (Å) | Si: 0.910 ± 0.0045 | Si: 1.152 ± 0.039 |
| | | C: 0.928 ± 0.017 | C: 1.30 ± 0.021 |
|  | Topmost SiC occupancy (atom/$A_{UC}$) | Si: 0.914 ± 0.014<br>C: 0.967 ± 0.027 | Si: 0.350 ± 0.046<br>C: 1.998 ± 0.076 |
|  | Bulk water height (Å) | 5.489 ± 0.0095 | 6.970 ± 1.88 |
|  | Bulk water peak rms width (Å) | 0.299 ± 0.017 | 0.814 ± 1.78 |
|  | Distance between layered water peaks (Å) | 3.158 ± 0.032 | 0.202 ± 0.233 |
|  | $\chi^2$ | 1.326 | 3.151 |
| 2 layers of graphene with partial coverage, bulk water | Graphene layer occupancy (C/$A_{UC}$) | $G_0$: 3.237 ± 0.087<br>$G_1$: 3.150 ± 0.00068<br>$G_2$: 1.143 ± 0.062 | $G_0$: 3.273 ± 0.035<br>$G_1$: 3.350 ± 0.064<br>$G_2$: 1.113 ± 0.064 |
|  | Graphene layer rms width (Å) | $G_0$: 0.300 ± 0.015<br>$G_{n>0}$: 0.139 ± 0.0080 | $G_0$: 0.343 ± 0.011<br>$G_{n>0}$: 0.169 ± 0.0084 |
|  | Graphene $d$ spacing (Å) | 3.408 ± 0.0077 | 3.407 ± 0.010 |
|  | Topmost SiC position (Å) | Si: 0.982 ± 0.014<br>C: 1.057 ± 0.041 | Si: 0.994 ± 0.012<br>C: 1.111 ± 0.026 |
|  | Topmost SiC occupancy (atom/$A_{UC}$) | Si: 0.840 ± 0.047<br>C: 1.297 ± 0.096 | Si: 0.732 ± 0.038<br>C: 1.354 ± 0.093 |
|  | Bulk water height (Å) | 2.549 ± 0.72 | 3.008 ± 0.15 |
|  | Bulk water peak rms width (Å) | 2.672 ± 0.75 | 1.498 ± 0.12 |
|  | Distance between layered water peaks (Å) | 5.406 ± 0.96 | 4.952 ± 0.089 |
|  | $\chi^2$ | 4.293 | 2.182 |

*Crystal truncation rod results over applied potential*

We collected crystal truncation rod (CTR) data on both graphene in water and graphene in 1 mM $YCl_3$, the latter of which was considered at different applied potentials. CTR signal for graphene in 1 mM $YCl_3$ at -0.6, -0.5, and 0.6 V (Figure S4) did not differ significantly, meaning the interfacial structure accessible via typical non-resonant CTR does not vary with applied potential. Therefore, we consider CTR data collected for 10 mM $YCl_3$ at -0.5 V for further analysis.

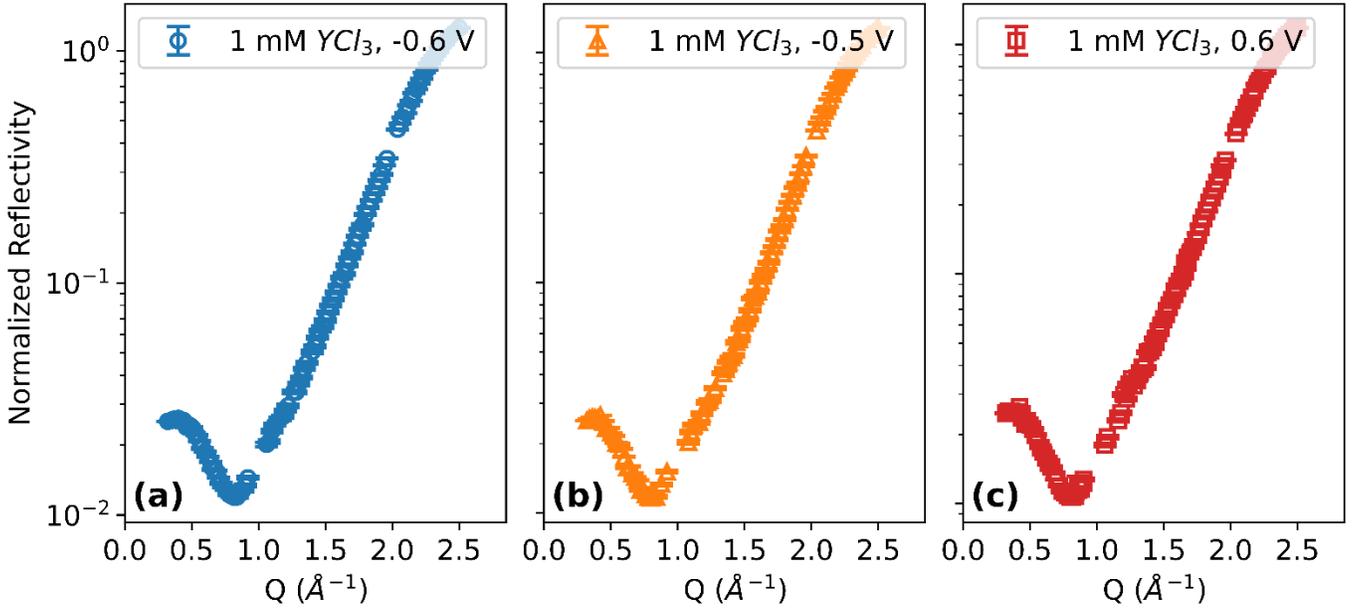

Figure S4. Crystal truncation rod reflectivity data for graphene in 1 mM YCl$_3$ at different applied voltages (panels) in the low $Q$ region. Data do not differ significantly across voltage.

*Resonant anomalous x-ray reflectivity fitting*

RAXR data can be described using the following total structure factor ($F$)

$$F(Q,E) = F_{NR}(Q) + F_R(Q,E)$$

$$= \sum_j O_j f_j(Q) \, exp(iQz_j) \, exp\left(\frac{-(Qu_j)^2}{2}\right) + [f'_Y(E) + if''_Y(E)]F_Y(Q) \qquad (S4)$$

The first term of Equation S4 is the non-resonant (NR) structure factor and describes the contribution of all $j$ atoms, including Y, i.e., the structure factor derived in the CTR measurements. The second term is the resonant (R) structure factor, which is composed of the resonant anomalous dispersion component, $f'_Y(E) + f''_Y(E)$, and partial structure factor ($F_Y(Q)$) of resonant atom Y. Chemical sensitivity comes from the resonant anomalous dispersion component, $f'_Y(E) + f''_Y(E)$, which we compute by measuring the x-ray absorption spectroscopy of a standard solution (100 mM YCl$_3$) in transmission mode and applying the Kramers-Kronig transform.[5] The $F_Y(Q)$ was first derived using a model-independent method via[6, 7]

$$F_Y(Q,E) = A_R(Q) exp(iP_R(Q)) \qquad (S5)$$

where $A_R$ and $P_R$ are its amplitude and phase at specific $Q$. At $Q << 1$, the $A_R$ and $P_R/Q$ can be used to describe the resonant atom occupancy and average position, respectively. We fitted our obtained RAXR spectra to Equation S4 using the least sum of squares and determined $A_R$ and $P_R$.

From these fits, we determined an appropriate structural model to describe Y adsorption at the graphene interface. We considered several models with varying the number of adsorbed layers, expressed as Gaussian peaks having different locations and coverages, combined with a diffuse distribution of yttrium beyond the surface, expressed as an exponential decay profile. The partial structure factor of the diffuse yttrium layer, $F_{Diff}$, was calculated using[8]

$$F_{Diff} = \frac{F_{Diff,1}}{1 - e^{(iQ - \kappa^{-1})d_{Diff}}} \tag{S6}$$

where $F_{Diff,1}$ is the structural factor of Gaussian peak used to represent the first Y layer in the decay profile, $\kappa^{-1}$ is the Debye length, and $d_{Diff}$ is the distance between the two adjacent Y layers within the profile. The best-fit model, having one adsorbed layer and a diffuse layer of yttrium, was selected the most appropriate model via $\chi^2$.

*Model-dependent resonant anomalous x-ray reflectivity fitting results over Debye length*

We fit our resonant anomalous x-ray reflectivity (RAXR) data for graphene in 10 mM $YCl_3$ using a structural model that includes an inputted Debye length. The theoretical Debye length for this system is 12.4 Å. We also fitted the data with different Debye lengths (Figure S5) and found similar adsorbed $Y^{3+}$ distributions. The largest difference between the fits is the total ion coverage that is positively correlated with the Debye length applied as a fixed input parameter. Using a Debye length of 25 Å, the occupancy of the diffuse layer is 1.16 ± 0.13 and the $\chi^2$ value is 2.69. A Debye length of 50 Å yields a diffuse layer occupancy of 2.143 ± 0.23 and a $\chi^2$ value of 2.64. We conclude the Debye length is not uniquely defined within our fitting process, as changing the Debye length yields similar results given the length of the diffuse layer.

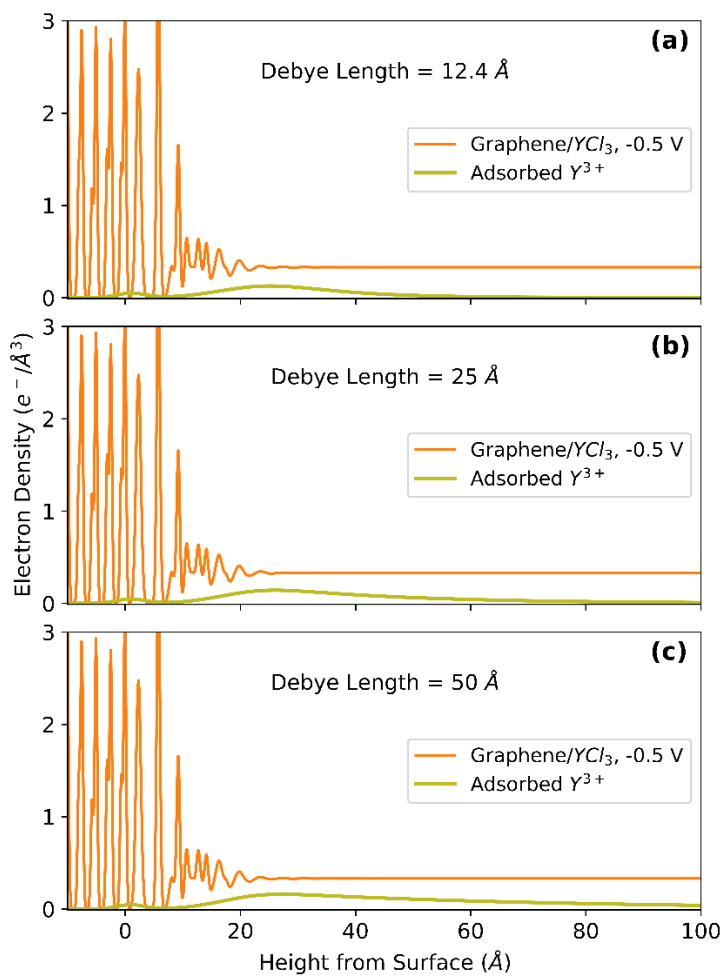

Figure S5. Total electron density profile derived from CTR analysis for graphene in 10 mM $YCl_3$ (orange) plotted with the Y electron density profile derived from RAXR (yellow) as a function of height from the SiC surface. Different Debye lengths are tested within the modelled non-resonant structure calculation (panels) and yield similar fits.